\documentclass[12pt,fleqn]{article}
\renewcommand{\baselinestretch}{1.0}

\usepackage{graphicx}

\begin{document}

\title{\Large Thermodynamic curvature and phase transitions in Kerr-Newman black holes}

\author{
   George Ruppeiner\footnote{ruppeiner@ncf.edu}\\
   Division of Natural Sciences\\
   New College of Florida\\
   5800 Bay Shore Road\\
   Sarasota, Florida 34243-2109 }

\maketitle

\begin{abstract}

Singularities in the thermodynamics of Kerr-Newman black holes are commonly associated with phase transitions.  However, such interpretations are complicated by a lack of stability and, more significantly, by a lack of conclusive insight from microscopic models.  Here, I focus on the later problem.  I use the thermodynamic Riemannian curvature scalar $R$ as a try to get microscopic information from the known thermodynamics.  The hope is that this could facilitate matching black hole thermodynamics to known models of statistical mechanics.  For the Kerr-Newman black hole, the sign of $R$ is mostly positive, in contrast to that for ordinary thermodynamic models, where $R$ is mostly negative.  Cases with negative $R$ include most of the simple critical point models.  An exception is the Fermi gas, which has positive $R$.  I demonstrate several exact correspondences between the two-dimensional Fermi gas and the extremal Kerr-Newman black hole.  Away from the extremal case, $R$ diverges to $+\infty$ along curves of diverging heat capacities $C_{J,\Phi}$ and $C_{\Omega,Q}$, but not along the Davies curve of diverging $C_{J,Q}$.  Finding statistical mechanical models with like behavior might yield additional insight into the microscopic properties of black holes.  I also discuss a possible physical interpretation of $|R|$.

\end{abstract}

\noindent
{\bf Suggested PACS Numbers}: 04.60.-m, 04.70.Dy, 05.40.-a

\section{INTRODUCTION}

\par
A Kerr-Newman black hole is characterized solely by its mass $M$, angular momentum $J$, and charge $Q$ \cite{Grav}.  Such simplicity allows a thermodynamic representation with laws analogous to the standard laws of thermodynamics \cite{Bek, Hawk, Dav, Hut}.  Previously, I discussed this structure in the context of thermodynamic fluctuation theory \cite{Rupp07}.  This leads naturally to thermodynamic Riemannian geometry; see \cite{Rupp95} for a review.

\par
The resulting thermodynamic Riemannian curvature scalar $R$ has been explored by a number of authors for black hole thermodynamics \cite{Ferrara,Cai99, Aman03, John, Arcioni, Shen, Aman2, Aman3, Aman4, Sarkar, Mirza, Quevedo, Aman5, Quevedo2, Aman6, Quevedo3,Medved,Myung}.  The main contribution of the present paper is an attempt at a physical interpretation of $R$, and its systematic evaluation for the Kerr-Newman black hole.  The analogy with ordinary thermodynamics is emphasized, as is the significance of the sign of $R$.

\par
The thermodynamic fluctuation formalism requires stability, namely, fluctuations about a maximum in the total entropy.  This issue poses difficulty for black holes.  In \cite{Rupp07} stability was obtained by restricting the number of independent fluctuating variables.  In addition, an infinite extensive environment was employed to have the fluctuations depend only on the known thermodynamics of the black hole.

\par
For an ordinary thermodynamic system, $|R|$ was interpreted \cite{Rupp79} as proportional to the correlation volume $\xi^d$, where $d$ is the system's spatial dimensionality and $\xi$ is its correlation length.  Direct calculations in a number of statistical mechanical models have verified this \cite{Rupp95}; see \cite{John} for a more recent review.  A thermodynamic quantity, $R$, then reveals information normally thought to reside in the microscopic regime, $\xi$.  Thus, $R$ has been of interest in black hole physics, which has good thermodynamic structures, but little conclusive microscopic information.

\par
I interpret $|R|$ for black holes as the average number of correlated Planck areas on the event horizon.  Although I give no direct microscopic model evidence, this interpretation would seem to be well motivated by the analogy with ordinary thermodynamics.\footnote{Such an interpretation is consistent with the assumption "that all the statistical degrees of freedom of a black hole live {\it on the black hole event horizon}" \cite{Park}.}  Zero $R$ indicates then "pixels" or "bits"\footnote{See \cite{BekSci} for a semipopular discussion picturing the quantization of area on the event horizon in terms of Planck areas.} on the event horizon fluctuating independently of each other.  Diverging $|R|$, which I take as signalling  a phase transition, indicates highly correlated pixels.

\par
$R$ diverges for extremal Kerr-Newman black holes, where the temperature  $T\rightarrow 0$.  I demonstrate here, and previously \cite{Rupp07b}, several exact limiting results matching extremal Kerr-Newman black hole thermodynamics to the two-dimensional (2D) Fermi gas ($d=2$).  Two dimensions are consistent with the membrane paradigm of black holes \cite{Thorne}.

\par
I also find instances of diverging $R$ where the heat capacities $C_{J,\Phi}$ and $C_{\Omega,Q}$ diverge and change sign, signalling a change of stability.  Although divergences in these heat capacities were identified by Tranah and Landsberg \cite{Lands} in 1980, they have been little discussed in the literature.  In contrast, I find no diverging $R$ along the Davies curve \cite{Dav} where the more familiar heat capacity $C_{J,Q}$ diverges.

\par
This paper is organized as follows.  First, I review the thermodynamic fluctuation picture in \cite{Rupp07}.  Second, I discuss thermodynamic Riemannian geometry and curvature, including an attempt at a physical interpretation for $R$.  Third, I calculate $R$ for the Kerr-Newman black hole.  Fourth, I compare the results with those in ordinary thermodynamics.

\section{THERMODYNAMIC FLUCTUATION THEORY}

\par
A major element in my approach is that the black hole resides in an infinite environment, characterized by mass $M_e$, angular momentum $J_e$, and charge $Q_e$.  The thermodynamics of the environment should be extensive; namely, $M_e$, $J_e$, and $Q_e$ should each scale up in proportion to the environment's volume.  With this structure, thermodynamic fluctuations require only the known thermodynamics of the black hole.  The environment's thermodynamic properties, which might be difficult to determine (dark matter, etc.), only sets the state about which fluctuations occur.  

\par
This structure is thermodynamically unstable if we allow an exchange of all three variables $(M,J,Q)$ \cite{Rupp07,Lands}.  Stability requires either a finite environment or a restriction on the number of independent fluctuating variables.  In \cite{Rupp07} I took the later approach, and considered the stability of seven cases in an infinite environment: fluctuating $(M,J,Q)$, $(J,Q)$, $(M,Q)$, $(M,J)$, $M$, $J$, and $Q$.\footnote{Kaburaki {\it et al.} \cite{Kabu93} considered the stability of the Kerr black hole ($Q=0$) under a variety of conditions, including ones similar to those here.  These authors used the sophisticated Poincar$\acute{\mbox{e}}$ turning point method which allows stability statements for finite environments with thermodynamics not necessarily known.  With an infinite environment, however, the black hole entropy Hessian determinants are sufficient for considerations of stability \cite{Rupp07}.  Fluctuating conserved variables $M$, $J$, or $Q$ are commonly referred to as "canonical," and fixed conserved variables as "microcanonical."}  Physically, we imagine that one (or two) of $M$, $J$, or $Q$ is so slow fluctuating that we can consider it to be essentially fixed.

\par
I use geometrized units with $M$ and $Q$ in cm, and  $J$ and entropy $S$ in cm$^2$ \cite{Grav}.  Useful are the Planck length

\begin{equation} L_p \equiv \sqrt{\frac{\hbar G}{c^3}} = 1.616\times 10^{-33} \mbox{cm},\label{480}\end{equation}

\noindent and the Planck mass

\begin{equation} M_p \equiv \sqrt{\frac{\hbar c}{G}}=2.177\times 10^{-5} \mbox{g},\label{490}\end{equation}

\noindent with $\hbar$, $c$, and $G$ the usual physical constants.  In geometrized units $G=c=1$ and $L_p=M_p$.

\par
The entropy of the Kerr-Newman black hole is \cite{Dav, Smarr}

\begin{equation} S(M,J,Q)=\frac{1}{8}\left(2M^2-Q^2+2\sqrt{M^4-J^2-M^2 Q^2}\right).\label{470} \end{equation}

\noindent To convert $S$ to real units, where it is $S_{bh}$, use

\begin{equation} \frac{S_{bh}}{k_B} = \left(\frac{8\pi}{L_p^2}\right)S,\label{500}\end{equation}

\noindent with $k_B$ Boltzmann's constant  \cite{Rupp07}.  

\par The total entropy of the universe is

\begin{equation} S_{tot}=S_{bh}+S_e,\label{510} \end{equation}

\noindent where $S_e$ is the entropy of the black hole's environment.  The fluctuation probability is given by Einstein's formula \cite{Landau},

\begin{equation} P\propto\mbox{exp}\left(\frac{S_{tot}}{k_B}\right).\label{520}\end{equation}

\par
Introduce the notation

\begin{equation} (X^1,X^2,X^3) \equiv (M,J,Q)  \end{equation}

\noindent and

\begin{equation} F_\alpha \equiv \frac{\partial S_{bh}}{\partial X^\alpha},\label{521}\end{equation}

\noindent with corresponding properties of the environment denoted by the subscript $e$.  The intensive $F_{e\alpha}$ values are independent of the size of the environment.

\par
Let us assume (incorrectly, as it turns out) that the black hole and the environment are fully in equilibrium, with a local maximum for $S_{tot}$.  Consider a small fluctuation $\Delta X^\alpha$ away from this equilibrium.  Expanding each of the entropies in Eq. (\ref{510}) to second order yields

\begin{equation} \Delta S_{tot} = F_\mu \Delta X^\mu + F_{e\mu} \Delta X_e^\mu + \frac{1}{2}\frac{\partial F_\mu}{\partial X^\nu}\Delta X^\mu \Delta X^\nu +
\frac{1}{2}\frac{\partial F_{e\mu}}{\partial X_e^\nu}\Delta X_e^\mu\Delta X_e^\nu, \label{522}\end{equation}

\noindent where the coefficients are evaluated at the equilibrium state, which is set by the environment.  The conservation laws demand

\begin{equation} \Delta X^{\alpha}=-\Delta X_e^{\alpha}, \label{523}\end{equation}

\noindent and a necessary condition for maximum entropy is

\begin{equation} F_\alpha = F_{e\alpha}.\label{524}\end{equation}

\par
With a very large environment, the second quadratic term in Eq. (\ref{522}) is negligible compared with the first.  To see this, fix the values of $\Delta X_e^\alpha$, which equal $-\Delta X^\alpha$.  As the environment is scaled up to infinite size at fixed $F_{e\alpha}$, $X_e^\alpha$ scales up in proportion without limit, and $\partial F_{e\alpha}/\partial X_e^\beta\rightarrow 0$.  The ability to drop this second quadratic term is a significant simplification offered by an infinite, extensive environment.

\par Eq. (\ref{522}) now can be written as

\begin{equation} \frac{\Delta S_{tot}}{k_B} =-\frac{1}{2}g_{\mu\nu}\Delta X^\mu\Delta X^\nu, \label{525}\end{equation}

\noindent where the symmetric matrix\footnote{In \cite{Rupp07}, the symbol $\gamma_{\alpha\beta}$ was used for this quantity.  $g_{\alpha\beta}$ denoted this quantity without the unit conversion factor for $S$.}

\begin{equation} g_{\alpha\beta} \equiv -\left(\frac{8\pi}{L_p^2}\right)\frac{\partial^2 S}{\partial X^\alpha\partial X^\beta}. \label{560}\end{equation}

\par
The Gaussian approximation to the fluctuation probability is

\begin{equation} P dX^1dX^2dX^n = \frac{\sqrt{|g|}}{(2\pi)^{\frac{n}{2}}}\mbox{ exp}\left(-\frac{1}{2}g_{\mu\nu}\Delta X^\mu \Delta X^\nu\right)dX^1dX^2dX^n, \label{570}\end{equation}

\noindent where $|g|$ is the determinant of $g_{\alpha\beta}$ and $n=3$ is the number of independent fluctuating variables.  If we set one or two $\Delta X^\alpha$'s to zero, reducing the value of $n$, Eqs. (\ref{525}) and (\ref{570}) are only trivially modified.  Entropy maximum requires that the matrix $g_{\alpha\beta}$ of the remaining variable(s) be positive definite.  Complete discussion of this is given in \cite{Rupp07}.

\par
The first fluctuation moments are zero \cite{Landau}:

\begin{equation} \left<\Delta X^\alpha\right> = 0.\label{580}\end{equation}

\noindent The second fluctuation moments are 

\begin{equation} \left<\Delta X^\alpha\Delta X^\beta\right> = g^{\alpha\beta}, \label{590}\end{equation}

\noindent with $g^{\alpha\beta}$ the components of the inverse $g_{\alpha\beta}$ matrix.

\par
Further notation is given in the Appendix, where I define the basic thermodynamic variables $T$, $\Omega$, and $\Phi$, the simplifying variables $\alpha$, $\beta$, $K$, and $L$, the entropy Hessian determinants $p_2$, $p'_2$, and $p''_2$, with numerators $A$, $B$, and $C$, and the heat capacities $C_{J,Q}$, $C_{J,\Phi}$, and $C_{\Omega,Q}$.  Diverging heat capacities are important below, and Figure 1 shows curves of infinities as well as the extremal limiting curve where the temperature $T\rightarrow 0$.

\section{THERMODYNAMIC RIEMANNIAN GEOMETRY}

\par
In this section, I summarize the thermodynamic Riemannian geometry.

\subsection{Thermodynamic metric}

\par
The metric follows naturally from the observation that the quadratic form in Eq. (\ref{525}) transforms as a scalar under any coordinate change because $\Delta S_{tot}$ depends only on the initial and final thermodynamic states.  Hence,

\begin{equation} (\Delta l)^2 = -\frac{2\Delta S_{tot}}{k_B} = g_{\mu\nu}\Delta X^\mu\Delta X^\nu \label{600}\end{equation}

\noindent
is a Riemannian line element.  It is unitless and positive definite assuming stability.  Its physical interpretation is clear from Eq. (\ref{570}): {\it the less probable a fluctuation between two states, the further apart they are.}

\par
In the definition of $g_{\alpha\beta}$ in Eq. (\ref{560}), $S$ was converted to $S_{bh}/k_B$ in real units, essential in Eq. (\ref{520}).  Such a unit conversion is unnecessary if $R$ is not needed beyond a proportionality constant.  However, a quantitative interpretation of $R$ in analogy with ordinary thermodynamics requires a unitless line element of the form in the exponential of Eq. (\ref{570}).

\par
To get the metric elements in Eq. (\ref{560}), I used the special properties of the conserved variables $(M,J,Q)$.  Once we have Eq. (\ref{560}), $g_{\alpha\beta}$ transforms as a second rank tensor under a change of coordinates \cite{Rupp95}.  Generally, under such a transformation the Hessian form in Eq. (\ref{560}) will not persist.  However, since we know the function $S=S(M,J,Q)$, there is no need in this paper to change coordinates.

\subsection{Thermodynamic curvature}

\par
Calculate $R$ as follows \cite{Weinberg}: the Christoffel symbols are

\begin{equation} \Gamma^{\alpha}_{\beta\gamma}=\frac{1}{2}g^{\mu\alpha}\left(g_{\mu\beta,\gamma}+g_{\mu\gamma,\beta}-g{_{\beta\gamma,\mu}} \right),\label{610}\end{equation}

\noindent where the comma notation indicates partial differentiation.  The Riemannian curvature tensor is

\begin{equation}  R^{\alpha}_{\beta\gamma\delta}=\Gamma^{\alpha}_{\beta\gamma,\delta}-\Gamma^{\alpha}_{\beta\delta,\gamma}+
\Gamma^{\mu}_{\beta\gamma}\Gamma^{\alpha}_{\mu\delta}-\Gamma^{\mu}_{\beta\delta}\Gamma^{\alpha}_{\mu\gamma},\label{620}\end{equation}

\noindent and the Riemannian curvature scalar is

\begin{equation} R=g^{\mu\nu}R^{\xi}_{\mu\xi\nu}.\label{630}\end{equation}

\noindent $R$ is independent of the choice of coordinate system, suggesting it is a fundamental measure of thermodynamic properties.  Since the line element is unitless, $R$ will be unitless.

\par
For two-dimensional thermodynamic geometries ($n=2$), all components of the Riemannian curvature tensor may be expressed in terms of the curvature scalar $R$ \cite{Weinberg}.  Not so in higher dimensions.  However, it was argued  \cite{Rupp98} that $R$ is the essential quantity in thermodynamic geometry regardless of the number of independent thermodynamic variables.

\par
For an ordinary pure fluid, a common picture \cite{Rupp79} is that of an open subsystem with fixed volume $V$ surrounded by an infinite environment of the same fluid.  The entropy fluctuation is

\begin{equation} \frac{\Delta S_{tot}}{k_B} = \frac{1}{2}V\frac{1}{k_B}\frac{\partial^2 s}{\partial x^\mu\partial x^\nu}\Delta x^\mu\Delta x^\nu, \label{640}\end{equation}

\noindent where $s$ is the entropy per volume (in units of $k_B$ per volume), and $x^1$ and $x^2$ are the energy and particle number per volume, respectively.  The pure fluid line element was written \cite{Rupp79} with $V$ omitted:

\begin{equation} (\Delta l)^2 = -\frac{1}{k_B}\frac{\partial^2 s}{\partial x^\mu\partial x^\nu}\Delta x^\mu\Delta x^\nu, \label{650} \end{equation}

\noindent and has units of inverse volume.  The corresponding $R$ has units of volume.

\par
Logically, however, the pure fluid line element could have been written in the unitless form

\begin{equation} (\Delta l)^2 = -\frac{2\Delta S_{tot}}{k_B}=-\frac{1}{k_B}\frac{\partial^2 S}{\partial X^\mu\partial X^\nu}\Delta X^\mu\Delta X^\nu, \label{660}\end{equation}

\noindent with neither the subsystem entropy $S$ nor the conserved energy and particle number $\{X^1$, $X^2\}$ divided by the constant $V$.  The form of this line element matches that of the black hole line element Eq. (\ref{600}).  It has unitless $R$.

\par
For the pure fluid, it was noted  \cite{Rupp79} that $R$ calculated with the line element Eq. (\ref{650}) is zero for the pure ideal gas, suggesting that $R$ is a measure of intermolecular interaction strength.  Indeed, calculations showed $|R|$ to be proportional to the correlation volume $\xi^d$ for a number of statistical mechanical models \cite{Rupp95,John}.

\par
Such calculations dovetailed nicely with $R$ having units of volume.  However, the units of $R$ are not naturally determined.\footnote{My previous arguments about the significance of volume units for $R$ (see, e.g., Sec. VI.B of \cite{Rupp95}) may have been overstated.  Model calculations and the path integral approach to thermodynamic fluctuation theory \cite{Rupp83} are the best way to establish $R\propto\xi^d$.  However, the pulling out of $V$ in Eq. (\ref{650}), and the resulting units of volume for $R$, is certainly natural and leads to correct results.}  With the equally valid line element in Eq. (\ref{660}), the fixed $V$ now appears in the denominator of $R$, and $R$ is unitless.  If we imagine the fluid broken up into three-dimensional  ($d=3$) pieces each of volume $V$, $|R|$ is the average number of correlated "pixels."  The physical interpretation of $|R|$ is then essentially the same regardless of whether or not we pull $V$ out of the line element.

\par
This leads to a possibly useful way to look at black holes.  Although there is no fixed subsystem volume to set a scale, the Planck length $L_p$ suggests a physical constant for this role.\footnote{Note, in ordinary thermodynamics there is no physical constant with units of length.}  Black hole thermodynamics takes place on the 2D event horizon.  It is natural to break it up into square pixels each of area $L_p^2$ \cite{BekSci}.  By analogy with the pure fluid, I interpret $|R|$ as the average number of correlated pixels.  Figure 2 illustrates this physical interpretation.

\par
I cannot presently support this idea with microscopic calculations of a type which were so valuable in ordinary thermodynamics.  However, the correspondence in the extremal limit with the 2D Fermi model \cite{Rupp07b}, discussed below, indicates at least consistency with the black hole membrane paradigm \cite{Thorne} which puts all black hole properties on the 2D event horizon.

\par
Note, this interpretation is only of $|R|$.  Janyszek and Mruga{\l}a \cite{Mrug90} argued that the sign of $R$ is also important.  I amplify on this in Sec. 6.

\par
Finally, in a coordinate system with metric elements of Hessian form, $R$ simplifies.  The arguments in \cite{Rupp98} allow one to show that with metric elements in Eq. (\ref{560}),

\begin{equation}R=\frac{1}{4}g^{\mu\nu}g^{\xi o}g^{\pi\rho}\left(g_{\mu\nu,\xi}g_{o\pi,\rho}-g_{\mu o,\rho}g_{\nu\xi,\pi}\right).\end{equation}

\noindent The second derivatives of the metric elements cancel in the calculation. 

\subsection{Background on black hole thermodynamic curvature from the entropy metric}

\par
\r{A}man {\it et al.} \cite{Aman4} presented a recent review of thermodynamic curvature in the context of black holes, so my remarks in this section will be brief.  Ferrara {\it et al.} \cite{Ferrara} were the first to apply thermodynamic curvature to black holes, to calculate critical behavior in moduli spaces. Cai and Cho \cite{Cai99} connected phase transitions in Ba$\tilde{\mbox{n}}$ados-Teitelboim-Zanelli (BTZ) black holes to diverging $R$.  They also identified a correspondence with $R$ for the Takahashi gas, suggesting that an appropriate black hole statistical model might be a system of hard rods.

\par
 \r{A}man, Bengtsson, and Pidokrajt \cite{Aman03} were the first to evaluate $R$ for various instances of the Kerr-Newman black hole, especially the two-dimensional ($n=2$) Kerr and Reissner-Nordstr\"{o}m cases.  They also considered a nonzero cosmological constant.  Arcioni and Lozano-Tellechea \cite{Arcioni} worked out five-dimensional black holes and black rings, including an extensive review.  They connected phase transitions to both diverging $R$ and diverging second fluctuation moments.
 
 \par
 \r{A}man {\it et al.} \cite{Aman2} examined $R$ in the context of homogeneous functions, emphasizing in particular cases with $R=0$.  \r{A}man and Pidokrajt \cite{Aman3} investigated Kerr and Reissner-Nordstr\"{o}m black holes in spacetime dimensions higher than four.  They found that patterns in four dimensions continue to higher dimensions.  Sarkar {\it et al.} \cite{Sarkar} evaluated $R$ for a general class of BTZ black holes, including quantum corrections to the entropy.  Mirza and Zamaninasab \cite{Mirza} worked out the curvature of the full 3D geometry for the Kerr-Newman black hole.  They found that $R$ diverges at the extremal limit, but not along the Davies curve.  \r{A}man {\it et al.} \cite{Aman5} reported results on dilaton black holes.

\subsection{Curvature from other than entropy metrics}

One is certainly not constrained to do thermodynamic Riemannian geometry with fluctuations and its entropy metric.  Another possibility is to express the internal energy in terms of its natural variables, $M=M(S,J,Q)$, and construct an energy metric from its Hessian.  This was done originally in ordinary thermodynamics by Weinhold \cite{Weinhold}.  The energy metric is conformally equivalent to the entropy metric \cite{Salamon}, with the same angles between vectors, but different $R$.

\par
A motivation for exploring other metrics (including the energy metric) is a concern by some authors about the physical validity of cases with $R=0$ from the entropy metric.  If $R$ is interpreted as a measure of interactions among gravitating particles, one might logically expect $|R|$ to be uniformly large for black holes, where gravitational forces are very big.

\par
However, the interpretation of $R$ in this paper takes a different approach.  The gravitating particles have presumably collapsed to the central singularity, shrinking the interactions between them to zero volume.  The statistics underlying the thermodynamics are envisioned to be on the event horizon.  A result $R=0$ seems now not so unreasonable.  Yet, I present little in the way of microscopic evidence for this point of view, so concerns about the physical validity of $R=0$ certainly cannot be dismissed.

\par
For the Reissner-Nordstr\"{o}m black hole \r{A}man {\it et al.} \cite{Aman03} found $R=0$ with the entropy metric.  To avoid this zero curvature, Shen {\it et al.} \cite{Shen} constructed a new entropy metric, replacing $Q$ with $\Phi$ and $M$ with $M-\Phi Q$.  These authors also argued that $R$ should signal (by diverging) a phase transition at the Davies curve.  Their modified $R$ shows such a divergence.  A detailed analogy with the van der Waals phase transition was worked out with their modified metric.  The authors also connected to modern themes in particle theory, such as holography and the AdS/CFT correspondence.  Mirza and Zamaninasab \cite{Mirza} evaded zero curvature by evaluating $R$ for the full 3D Riemannian geometry.  Here, $R$ is always positive, as will be discussed in Sec. 5.1.  Quevedo and collaborators \cite{Quevedo, Quevedo2, Quevedo3} suggested that this issue requires Legendre invariant metrics to deal with properly.  They constructed a detailed framework based on this idea.  Medved \cite{Medved} also gave a recent discussion of these issues.

\section{Background on phase transitions} 

\par
Exactly what constitutes a black hole phase transition is somewhat unsettled in the literature.  In ordinary thermodynamics the modern belief is "a phase transition occurs when there is a singularity in the free energy or one of its derivatives" \cite{Yeomans}.  Phase transitions can bring about dramatic contrasts, like between a solid and a gas.  Or changes can be more subtle, like the onset of a gradual deformation in crystal structure.  Conjectured phase transitions in Kerr-Newman black holes are typically of the more subtle variety, second-order phase transitions associated perhaps with a diverging heat capacity.

\par
Phase transition theory in ordinary thermodynamics typically includes equilibrium between system and environment.  Achieving this with black holes can be difficult.  A more serious problem is the lack of conclusive microscopic models for black holes.  This makes it hard to identify objects as fundamental to phase transition theory as order parameters and correlation functions.

\par
There are then a number of viewpoints of what might be involved in a black hole phase transition: 1) a change in topology, 2) a divergence of a second fluctuation moment, 3) a divergence of a heat capacity, 4) an onset of instability like that in an axisymmetric rotating self-gravitating fluid, 5) a divergence of $R$, and 6) consistency with the scaling laws of critical phenomena.\footnote{Another possibility is the first-order phase transition consisting of a black hole condensing out of the background space \cite{Hut}.  However, this differs from the second-order phase transitions at issue here.}

\par
For the Kerr-Newman black hole, there is no topology change except possibly at the extremal limit.  Otherwise, the topology is that of the sphere \cite{Hawk2}.

\par
Second fluctuation moments are connected to quantities such as heat capacities through thermodynamic fluctuation theory \cite{Landau}, so viewpoints 2) and 3) are related, a point not always clear in the black hole literature.

\par
Davies \cite{Dav} argued that the curve of diverging $C_{J,Q}$ constitutes a second-order phase transition.  However, this has been disputed by a number of authors.  One issue is whether or not the Davies curve marks an actual change of stability.  I found that it does so only for $M$ fluctuations \cite{Rupp07}.  Davies  \cite{Dav} also brought up the analogy with the change of symmetry of an axisymmetric rotating self-gravitating fluid \cite{Bertin}.  However, this was questioned \cite{Sokol} since the nonrotating black hole also crosses the Davies curve as charge is increased.

\par
Arguments based on scaling theory usually involve attempts to introduce an order parameter.  Cai {\it et al.} \cite{Cai93} and Kaburaki \cite{Kabu96} argued that the extremal limit constitutes a second-order phase transition and proposed the difference between the inner and outer black hole radii as the order parameter.  Lousto \cite{Lousto93,Lousto97} emphasized instead a phase transition along the Davies curve.  He used $\Omega-\Omega_c$ as the order parameter, where $\Omega_c$ is the angular velocity along the Davies curve.  He also worked out critical exponents and discussed them in the context of scaling theory.  Lau \cite{Lau} also argued that the Davies curve corresponds to a second-order phase transition.

\section{KERR-NEWMAN THERMODYNAMIC CURVATURE}

\par
In this section, I work out $R$ for $(M,J,Q)$, $(J,Q)$, $(M,Q)$, and $(M,J)$ fluctuations.  $M$, $J$, and $Q$ fluctuations, with $n=1$, have $R$ trivially zero, and require no special consideration.\footnote{Trivial geometries ($n=1$) reflect noninteracting systems.  For example, open fluid systems characterized by one fluctuating parameter, usually the internal energy or the temperature, generally do not have interactions, e.g., a gas of photons.  With interactions, an additional parameter, like the density, is generally required.}

\par
I go beyond \r{A}man {\it et al.} \cite{Aman03}, and report all situations.  In the Kerr-Newman family, these authors focused primarily on Reissner-Nordstr\"{o}m black holes, represented by the geometry of $(M,Q)$ fluctuations with $J=0$, and Kerr black holes, represented by the geometry of $(M,J)$ fluctuations with $Q=0$.

\subsection{$(M, J, Q)$ fluctuating}

\par
Here, all $(M, J, Q)$ fluctuate.  By Eq. (\ref{600}),

\begin{equation} (\Delta l)^2 =\begin{array}{ll} g_{11}(\Delta M)^2 + 2 g_{12}\Delta M \Delta J + 2 g_{13}\Delta M \Delta Q+\\g_{22}(\Delta J)^2 + 2 g_{23}\Delta J\Delta Q + g_{33}(\Delta Q)^2, \end{array}\label{670}\end{equation}

\noindent corresponding to a 3D Riemannian geometry ($n=3$).  This case falls entirely outside the domain of stable fluctuations \cite{Rupp07,Lands}, and so I give it only a little attention.

\par
Evaluation with Eq. (\ref{630}) shows $R$ to be always real and positive, with a minimum of $(M_p/2\sqrt{\pi}M)^2$ at the origin $J=Q=0$.  $R$ is shown in Figure 3.\footnote{As was pointed out \cite{Aman03, Mirza}, the full algebraic expression for $R$ is not particularly revealing.  I will adhere to tradition and not show it here.}  It has no anomalies except at the extremal limit, where it diverges proportional to $T^{-1}$.

\par
Mirza and Zamaninasab \cite{Mirza} also worked out this case.  With zero cosmological constant, they found that $R$ diverges at the extremal limit, but  nowhere else.  In particular, they found no divergence along the Davies curve.  They also found nonzero $R$ for the Reissner-Nordstr\"{o}m case, $J=0$.  Fig. 3 corroborates these findings.

\subsection{$(J, Q)$ fluctuating, M fixed}

\par
Here, $(J,Q)$ fluctuate at fixed $M$.  By Eq. (\ref{600}),

\begin{equation} (\Delta l)^2 = g_{22}(\Delta J)^2 + 2 g_{23}\Delta J \Delta Q + g_{33}(\Delta Q)^2,\label{680} \end{equation}

\noindent corresponding to a 2D Riemannian geometry ($n=2$).  Fluctuations in this case are stable for all states in the physical regime \cite{Rupp07}.

\par
By Eq. (\ref{630}), and the definitions in the Appendix,

\begin{equation} R = \frac{\left(K^5+L^2 K^3-2 K^3-2 K^2+3 L^2 K-3 K+2\right)}{4\pi K B^2}\left(\frac{M_p}{M}\right)^2.\label{690}\end{equation}

\noindent $R$ is shown in Figure 4.  As is argued in the Appendix, $B$ is never zero in the physical regime.  Hence, $R$ only diverges at the extremal limit where $K\rightarrow 0$.

\par
Let us examine further the extremal limit.  Equations (\ref{690}), (\ref{890}), and (\ref{950}) yield the extremal limiting expressions

\begin{equation} C_{J,Q} =\frac {1}{16}M^3 L^2 T, \label{700}\end{equation}

\noindent and

\begin{equation}R =\frac{2 M_p^2}{\pi M^3 L^2 T}. \label{710}\end{equation}

\noindent The limiting product of curvature and heat capacity,

\begin{equation}\left(R\right)\left(\frac{8\pi}{L_p^2} C_{J,Q}\right)=\left(\frac{2 M_p^2}{\pi M^3 L^2 T}\right)\left(\frac{8\pi M^3 L^2 T}{16L_p^2}\right)=1,\label{720}\end{equation}

\noindent is a unitless, scale free constant independent of where we are on the extremal limiting curve.  In Section 6.2, I evaluate the statistical mechanics of the 2D Fermi gas and demonstrate several exact correspondences with these results at low temperature.

\par
We have from Eq. (\ref{590}), and the definitions in the Appendix, the dimensionless second fluctuation moments

\begin{equation} \frac{\sqrt{\left<(\Delta J)^2\right>}}{\hbar} = \frac{1}{\sqrt{2\pi}} \sqrt{\frac{K^4-L^2 K+2 K}{B}}\left(\frac{M}{M_p}\right),\label{730}\end{equation}

\noindent and

\begin{equation} \frac{\sqrt{\left<(\Delta Q)^2\right>}}{e}=\frac{1}{2}\sqrt{\frac{137.04}{\pi}} \sqrt{\frac{2(K^3+L^2 K-K)}{B}},\label{740}\end{equation}

\noindent where $e$ is the electron charge and $137.04=\hbar/e^2$ is the fine structure constant.  Both fluctuation moments are real and nondiverging in the entire physical regime.  They have maxima of $1/\sqrt{2\pi}$ and $3.302$, respectively, at the origin $J=Q=0$, and decrease to zero as $\sqrt{T}$ at the extremal limiting curve.

\subsection{$(M, Q)$ fluctuating, J fixed}

\par
Here, $(M,Q)$ fluctuate at fixed $J$.  By Eq. (\ref{600}),

\begin{equation} (\Delta l)^2 = g_{11}(\Delta M)^2 + 2 g_{13}\Delta M \Delta Q + g_{33}(\Delta Q)^2,\label{750}\end{equation}

\noindent corresponding to a 2D Riemannian geometry ($n=2$).  There is a slice of stability \cite{Rupp07} in the ($\sqrt{\alpha}$, $\sqrt{\beta}$) plane bounded by the extremal limiting curve and the curve $C=0$, as shown in Figure 5.

\par
By Eq. (\ref{630}), and the definitions in the Appendix,

\begin{equation} R = -\frac{(L-1) (L+1)}{2 \pi K C^2}\left( \begin{array}{llllll} 3 K L^6-4 L^6+4 K^3 L^4-\\8 K^2 L^4-6 K L^4+36 L^4+\\K^5 L^2-4 K^4 L^2+14 K^3 L^2+\\40 K^2 L^2-36 K L^2-96 L^2+\\8 K^5+4 K^4-36 K^3-\\32 K^2+48 K+64\end{array} \right)\left(\frac{M_p}{M}\right)^2.\label{760}\end{equation}

\noindent $R$ is shown in Figure 6. Despite only a limited slice of stability, located in the "saddlebags" near $J/M^2=\pm1$, $R$ is real everywhere in the physical regime.

\par
Along the line $J=0$, we clearly have $R=0$, since $L=1$.  This was demonstrated by \r{A}man {\it et al.} \cite{Aman03} who also pointed out that there is no curvature anomaly at the Davies point $(J/M^2,Q/M)=(0,0.8660)$.  I add that no point along the line $J=0$ lies in the stable slice, as is clear in Fig. 5.

\par
Figure 6 shows a steep drop to negative $R$ near $Q/M=\pm1$.  The cut away view shows this as a waterfall shape.  Such abrupt behavior, the only case of negative curvature for the Kerr-Newman black hole, reminds one of the abrupt change in sign for the Takahashi gas \cite{Rupp90B} and for the finite 1D Ising model \cite{Brody03}, which will be discussed further in Section 6.1.  None of these negative values fall into the stable slice, however.

\par
Closer to the extremal limit, $R$ comes up again, diverging to $+\infty$ at all points on the extremal limiting curve $K=0$ except $J=0$.  Equations (\ref{760}) and (\ref{890}) yield the extremal limiting expression

\begin{equation}R =\frac{2 M_p^2}{\pi M^3 L^2 T}.\label{773}\end{equation}

\noindent Despite the difference between $(M,Q)$ and $(J,Q)$ fluctuations, this limiting expression is the same as Eq. (\ref{710}) for $(J,Q)$ fluctuations, and the match to the 2D Fermi gas applies equally well here.

\par
$R$ has an additional divergence, to $+\infty$, at the other boundary of stability, $C=0$.  $R$ diverges as  $C^{-2}$.  $C_{J,\Phi}$ diverges as $C^{-1}$, by Eq. (\ref{970}).

\par
We have from Eq. (\ref{590}), and the definitions in the Appendix, the dimensionless second fluctuation moments

\begin{equation}\frac{\sqrt{\left<(\Delta M)^2\right>}}{m_e}=\frac{1}{\sqrt{2\pi}}\left(\frac{M_p}{m_e}\right)\sqrt{\frac{K^4-L^2 K+2 K}{C}},\label{780}\end{equation}

\noindent with $m_e$ the electron mass, and

\begin{equation} \frac{\sqrt{\left<(\Delta Q)^2\right>}}{e}=\frac{1}{2}\sqrt{\frac{137.04}{\pi}} \sqrt{\frac{-4 K^4+2 L^2 K^3-8 K^3+2 L^4 K}{C}}.\label{790}\end{equation}

\noindent Both these quantities are real in the slice of stability.  They go to zero as $\sqrt{T}$ at the extremal limit $K=0$, and diverge along the curve $C=0$.  Hence, changing stability is marked by both diverging $R$ and diverging fluctuations.

\par
Note, fluctuations in $M$ expressed in units of the electron mass are huge.  However, in units of the Planck mass they would be much smaller, on the order of the fluctuations in $J$ and $Q$.

\subsection{$(M, J)$ fluctuating, Q fixed}

Here, $(M,J)$ fluctuate at fixed $Q$.  By Eq. (\ref{600}),

\begin{equation} (\Delta l)^2 = g_{11}(\Delta M)^2 + 2 g_{12}\Delta M \Delta J + g_{22}(\Delta J)^2,\label{800}\end{equation}

\noindent corresponding to a 2D Riemannian geometry ($n=2$).  Stability is confined to a slice bounded by the extremal limiting curve and the $A=0$ curve, as shown in Figure 7.

\par
By Eq. (\ref{630}), and the definitions in the Appendix,

\begin{equation} R = \frac{1}{2\pi K A^2}      \left( \begin{array}{llll} K^7+3 K^6+2 L^2 K^5+6 L^2 K^4-\\5 K^4+L^4 K^3+9 L^2 K^3-9 K^3+\\3 L^4 K^2+4 L^2 K^2-8 K^2+9 L^4 K-\\21 L^2 K+12K+9L^4-24 L^2+16 \end{array} \right)\left(\frac{M_p}{M}\right)^2.\label{810}\end{equation}

\noindent It is shown in Figure 8.  Despite only a limited slice of stability, $R$ is real and positive everywhere in the physical regime.  Its minimum value is $(M_p/2\sqrt{\pi}M)^2$ at the origin.

\par
\r{A}man {\it et al.} \cite{Aman03} computed $R$ for $Q=0$, and found it to diverge at the extremal limit.  They pointed out that there is no curvature anomaly at the Davies point $(J/M^2,Q/M)=(0.6813,0)$.  This is confirmed by the findings here.  I add that no point with $Q=0$ lies in the stable regime, as is clear in Fig. 7.

\par $R$ diverges at the extremal limit $K=0$.  Equations (\ref{810}) and (\ref{890}) yield the extremal limiting expression

\begin{equation}R =\frac{2 M_p^2}{\pi M^3 L^2 T}.\label{775}\end{equation}

\noindent Remarkably, this is the same as Eqs. (\ref{710}) and (\ref{773}) found previously.

\par
$R$ has an additional divergence, to $+\infty$, at the other boundary of stability, $A=0$.  $R$ diverges as $A^{-2}$.  $C_{\Omega,Q}$ diverges as $A^{-1}$, by Eq. (\ref{990}). 

\par
We have from Eq. (\ref{590}), and the definitions in the Appendix, the dimensionless second fluctuation moments

\begin{equation}\frac{\sqrt{\left<(\Delta M)^2\right>}}{m_e}=\frac{1}{\sqrt{2\pi}}\left(\frac{M_p}{m_e}\right)\sqrt{\frac{K^3+L^2 K-K}{A}},\label{820}\end{equation}

\noindent and

\begin{equation} \frac{\sqrt{\left<(\Delta J)^2\right>}}{\hbar}=\frac{1}{\sqrt{2\pi}} \sqrt{\frac{-2 K^4+L^2 K^3-4 K^3+L^4 K}{A}}\left(\frac{M}{M_p}\right).\label{830}\end{equation}

\noindent Both these quantities are real in the slice of stability. They go to zero as $\sqrt{T}$ at the extremal limit $K=0$, and diverge along the curve $A=0$.  Hence, changing stability is marked by both diverging $R$ and diverging fluctuations.

\section{DISCUSSION}

In this section, I review evaluations of $R$ in ordinary thermodynamics, and compare with the Kerr-Newman black hole.

\subsection{Curvature in ordinary thermodynamic models}

\par
Table I reviews signs and divergences of  thermodynamic curvature in several ordinary thermodynamic models.  Most table entries are simple models where $R$ may be worked out in closed form.  The tendency is negative $R$ where attractive interactions dominate, and positive $R$ where repulsive interactions dominate.\footnote{One must guard against the impression that there is a connection between thermodynamic stability in the sense here and the sign of $R$.  It is tempting, for example, to envision the $n=2$ thermodynamic Riemannian geometry of the type here as a 2D surface embedded in a 3D flat Euclidean space from which it inherits its metric.  In such a construction, thermodynamic stability requires $R$ for the 2D embedded surface to be negative.  However, this picture is incorrect, as has been discussed in Sec. IV.G of Ref. \cite{Rupp95}.  There is no connection between thermodynamic stability and the sign of $R$.}  Janyszek and Mruga{\l}a \cite{Mrug90} emphasized the importance of the sign of $R$, and identified the quantum gasses, 3D Bose and 3D Fermi, as essential examples with opposite signs.

\par
The signs of $R$ for the standard critical point models in Table I are all negative, and have critical point divergences $R\rightarrow -\infty$.  This is quite unlike the Kerr-Newman black hole, with its predominantly positive $R$ and divergences $R\rightarrow+\infty$.

\par
Table I shows a group of weakly interacting systems with "small" $|R|$.  Small means on the order of the volume of an intermolecular spacing or less.  I view such values of $R$ as physically equivalent to zero, since the meaning of correlation volumes of this size is lost in the "noise" of thermodynamics breaking down as individual atoms and spins become visible.  The 1D antiferromagnetic Ising model  \cite{Rupp81,Mrug} is perhaps misplaced here, since its disaligning interactions might propagate a long way.  However, for antiferromagnets, the true ordering field is a staggered field, and not the constant field used for the calculations in Table I.  A reassessment of this model in these terms might be called for.

\par
There are three cases in Table I having both positive and negative curvatures.  The 1D $q$-state Potts model \cite{John,Dol02} has sign related to $q$.  For $q>2$, and nonzero field, there are significant regions of negative $R$ at low temperature.  The 2D Potts model has the dimensionality of the Kerr-Newman event horizon.  Its $R$ has not yet been evaluated, but perhaps its study could yield an appropriate critical line with positive $R$.

\par
The Takahashi gas \cite{Rupp90B} has the typical negative $R$ in the gaslike phase where attractive interactions dominate, and small $|R|$ in the liquidlike phase where interactions are short-range.  However, going from one phase to the other by changing the density at constant temperature, there is a pseudophase transition accompanied by a sharp {\it positive} curvature spike.  Cai and Cho \cite{Cai99} connected this spike to a phase transition in the BTZ black hole.

\par
An abrupt change in sign of $R$ is also present in the finite 1D Ising ferromagnet of $N$ spins \cite{Brody03}.  This model has the typical negative $R$ for large $N$, but a sharp rise to large positive values as $N$ is decreased.  Whether or not this result has relevance here is unclear.

\par
At the bottom of Table I there are the 3D Fermi gas \cite{Mrug90} and the 3D Fermi paramagnet \cite{Kav}.  Both models have positive $R$, diverging as $T\rightarrow 0$.  These results lead me now to take a closer look at Fermi gasses, particularly the 2D Fermi gas.

\subsection{Curvature for the 2D Fermi gas}

\par
For the 3D Fermi gas at low $T$, $R$ seems to diverge \cite{Mrug90} as $T^{-3/2}$, and not as $T^{-1}$ in Eq. (\ref{710}) for the Kerr-Newman black hole.  This motivates me to work out the 2D Fermi gas.  By the reasoning leading to Eq. (8.1.3) of \cite{Pathria}, the 2D Fermi gas has thermodynamic potential

\begin{equation} \phi(\frac{1}{T},-\frac{\mu}{T})=\frac{p}{T}=k_B g\lambda^{-2}f_2(\eta), \end{equation}

\noindent with pressure $p$, $\eta\equiv\mbox{exp}(\mu/k_B T)$, chemical potential $\mu$, thermal wavelength $\lambda\equiv h/\sqrt{2\pi m k_B T}$, particle mass $m$, weight factor $g\equiv(2s+1)$, particle spin $s$, and

\begin{equation} f_l(\eta)\equiv\frac{1}{\Gamma(l)}\int_0^\infty\frac{x^{l-1}dx}{\eta^{-1}e^x+1}. \label{gK}\end{equation}

\noindent I use obvious fluid units for all quantities, including $S$ and $T$.  The integral in Eq. (\ref{gK})  converges for $f_2(\eta)$, and yields $f_1(\eta)=\mbox{ln}(1+\eta)$.  $f_0(\eta)$ and $f_{-1}(\eta)$ follow from $f_1(\eta)$ using the recurrence relation $f_{l-1}(\eta)=\eta f'_l(\eta)$ \cite{Pathria}.

\par Define the heat capacity at constant particle number $N$ and constant area $A$ by

\begin{equation} C_{N,A} \equiv T\left(\frac{\partial S}{\partial T}\right)_{N,A}=N k_B\left[2 \frac{f_2(\eta)}{f_1(\eta)}-\frac{f_1(\eta)}{f_0(\eta)}\right].\end{equation}

\noindent The second equality is by Problem 8.10.ii of \cite{Pathria}.  The methods of \cite{Pathria} now yield the limiting low $T$ expression

\begin{equation} \frac{C_{N,A}}{Ak_B}=\frac{2\pi^3 g m k_B T}{3 h^2}. \label{gN}\end{equation}

\par Evaluating $R$ with Eq. (6.31) of \cite{Rupp95} yields

\begin{equation} R=-g^{-1}\lambda^2\left\{\frac{-2 f_2(\eta ) f_0(\eta )^2+f_1(\eta )^2 f_0(\eta )+f_{-1}(\eta ) f_1(\eta ) f_2(\eta )}{\left[f_1(\eta )^2-2 f_0(\eta ) f_2(\eta )\right]^2}\right\}.\end{equation}

\noindent Numerical evaluation over the physical range $-\infty<\mu<+\infty$ and $0<T<\infty$ indicates $R$ is always positive.  The methods of \cite{Pathria} yield the limiting low $T$ expression:

\begin{equation} R =\frac{3h^2}{2\pi^3 g m k_B T}. \label{gM}\end{equation}

\par The limiting $T$ dependences of  $C_{N,A}$ and $R$ match the corresponding Kerr-Newman black hole quantities Eqs. (\ref{700}) and (\ref{710}).  This connection to a 2D model is consistent with the membrane paradigm of black holes \cite{Thorne}.  Furthermore, the limiting product of curvature and heat capacity,

\begin{equation}\left(\frac{R}{A}\right)\left(\frac{C_{N,A}}{k_B}\right)=\left(\frac{3h^2}{2\pi^3 g m k_B T A}\right)\left(\frac{2\pi^3 g m k_B T A}{3 h^2}\right)=1,\label{gO} \end{equation}

\noindent is a unitless, scale free constant independent of density.  The factor $A$ below $R$ undoes the traditional pulling out of $A$ in the ordinary thermodynamic line element.  $R/A$ here is analogous to $R$ for the Kerr-Newman black hole.  The constant products Eqs. (\ref{720}) and (\ref{gO}) are equal, remarkable for systems apparently so different.

\par
However, note a key difference.  The Kerr-Newman black hole entropy Eq. (\ref{470}) does not go to zero in the extremal limit, as it does for the 2D Fermi gas with its unique ground state.  Resolution probably requires a more sophisticated Fermi gas model.  Note as well that I have presented no detailed correspondence between the Kerr-Newman black hole thermodynamics and a specific microscopic Fermi model.  Such a connection is necessary to make the results given here something more than a possibly useful direction to explore.

\subsection{Curvature for black hole critical points at nonzero $T$}

\par
For the phase transitions found at the non-extremal boundaries no appropriate models with evaluated $R$'s present themselves.  The signs of $R$ of the simple critical point models in Table I are all negative, in contrast to the Kerr-Newman black hole results.  Hence, I make no attempt here  to suggest an order parameter or to  evaluate and interpret possible critical exponents and scaling relations between them.

\section{CONCLUSIONS}

In conclusion, the following were done in this paper for thermodynamic Riemannian geometry based on the entropy metric.

\par
First, I attempted a physical interpretation of $R$ for black holes.  It was based on analogy with the interpretation in ordinary thermodynamics.  Perhaps, this interpretation lessens concern over the physical plausibility of the occasional result $R=0$.

\par
Second, I reviewed previous evaluations of $R$ and phase transitions in Kerr-Newman black holes.

\par
Third, I gave a complete evaluation of $R$ for the Kerr-Newman black hole.  In all cases, $R$ was found to be positive in stable fluctuation regimes and to diverge to $+\infty$ at the extremal limit.  I also found $R$ to diverge to $+\infty$ at nonzero temperatures along curves of changing stability, where the heat capacities $C_{J,\Phi}$ and $C_{\Omega,Q}$ diverge.

\par
Fourth, I argued that the sign of $R$ is important, and tabulated signs in a number of ordinary thermodynamic models.  I found that most of the simple critical point models have negative $R$.  This might make them problematic for understanding Kerr-Newman black hole phase transitions.  Different models might be required.

\par
Fifth, I noted that the Fermi gas is one of the few known cases in ordinary thermodynamics with large positive $R$.  I established several exact correspondences between the 2D Fermi gas and the extremal Kerr-Newman black hole.  This suggests that microscopic models with fermions might be useful as a framework for formulating a microscopic description of black holes.

\section{APPENDIX: NOTATION}

\par
Notation was defined in \cite{Rupp07}, and is summarized here.  I differ only with the metric elements $g_{\alpha\beta}$, including here the unit conversion factor for $S$ in Eq. (\ref{500}).

\par
Define the temperature $T$, the angular velocity $\Omega$, and the electric potential $\Phi$ by \cite{Dav,Lands}

\begin{equation} \frac{1}{T} \equiv \left(\frac{\partial S}{\partial M}\right)_{J,Q}, \label{840}\end{equation}

\begin{equation} -\frac{\Omega}{T} \equiv \left(\frac{\partial S}{\partial J}\right)_{M,Q}, \label{850}\end{equation}

\noindent and

\begin{equation} -\frac{\Phi}{T} \equiv \left(\frac{\partial S}{\partial Q}\right)_{M,J}.\label{860} \end{equation}

\noindent Two standard unitless variables are \cite{Dav}

\begin{equation} \{\alpha,\beta\} \equiv\{J^2/M^4,Q^2/M^2\}. \label{870}\end{equation}

\noindent Simplifying the notation are \cite{Lands}

\begin{equation} \{K,L\}\equiv\{\sqrt{1-\alpha-\beta},\sqrt{1+\alpha}\}.\label{880}\end{equation}

\noindent We may show that

\begin{equation}\frac{1}{T}=\frac{\left(K^2+2 K+L^2\right) M}{4 K}.\label{890}\end{equation}

\noindent To be in the {\it physical regime} of real $S$ and $T$ requires

\begin{equation} \alpha + \beta < 1.  \label{900}\end{equation}

\noindent The curve of equality, $\alpha+ \beta=1$, has $K=T=0$ and constitutes the {\it extremal limit}, thought to be unattainable by the third law of black hole thermodynamics \cite{Carter}.

\par Major components in the discussion of stability are the entropy Hessian determinants:\footnote{In \cite{Rupp07} the metric elements $g_{\alpha\beta}$ did not include the conversion factor for $S$, we must undo these to make the entropy Hessian determinants the same as in \cite{Rupp07}.}

\begin{equation} p_2 \equiv \left(\frac{L_p^2}{8\pi}\right)^2\left| \begin{array}{cc} g_{11}&g_{12}\\g_{21}&g_{22} \end{array} \right|=\frac{-2 K^3-3 K^2-2 L^2 K+2 K-3 L^2+4}{16 K^4 M^2},\label{920}\end{equation}

\begin{equation} p'_2\equiv \left(\frac{L_p^2}{8\pi}\right)^2 \left| \begin{array}{cc} g_{22}&g_{23}\\g_{32}&g_{33}  \end{array} \right|=\frac{K^3+L^2 K-K+1}{16 M^2 K^4}, \label{930}\end{equation}

\noindent and

\begin{equation} p''_2 \equiv \left(\frac{L_p^2}{8\pi}\right)^2\left| \begin{array}{cc} g_{11}&g_{13}\\g_{31}&g_{33} \end{array}\right|=\frac{1}{16 K^4} \left(\begin{array}{llll}-K^4+L^2 K^3-4 K^3-\\L^2 K^2-2 K^2+L^4 K+\\2 L^2 K-4 K-2 L^4+\\10 L^2-8 \end{array}\right).  \label{940}\end{equation}

\noindent The numerators of $p_2$, $p'_2$, and $p''_2$ are, respectively,

\begin{equation} A=-2 K^3-3 K^2-2 L^2 K+2 K-3 L^2+4,\label{921}\end{equation}

\begin{equation} B=K^3+L^2 K-K+1,\label{922}\end{equation}

\noindent and

\begin{equation} C=\left(\begin{array}{ll}-K^4+L^2 K^3-4 K^3-L^2 K^2-2 K^2+\\L^4 K+2 L^2 K-4 K-2 L^4+10 L^2-8 \end{array}\right).\label{923}\end{equation}

\par
Curves along which these numerators go to zero identify changes of stability accompanied by divergences of heat capacities.  $A=0$ in the physical regime if and only if

\begin{equation} \alpha =  \frac{(3-4\beta)\beta^2}{4(\beta-1)^2}. \end{equation}

\noindent This curve is shown in Fig. 7.  $B$ is never zero in the physical regime, since $K\geq0$ and $L\geq 1$.  $C=0$ along a single curve in the physical regime, shown in Fig. 5, with its algebraic expression too complicated to show here.

\par
Finally, the heat capacities \cite{Lands}

\begin{equation} C_{J,Q} \equiv T\left(\frac{\partial S}{\partial T}\right)_{J,Q}=\frac {M^2 K(K^2+L^2+2K)}{4(L^2-2K)}, \label{950}\end{equation}

\begin{equation}  C_{J,\Phi} \equiv T\left(\frac{\partial S}{\partial T}\right)_{J,\Phi},\label{960}\end{equation}

\noindent which evaluates to

\begin{equation}  C_{J,\Phi}=\frac{M^2 K \left(K^2+2 K+L^2\right)}{4 C}\left(\begin{array}{ll} -L^4-K^2 L^2+K L^2+4 L^2+\\K^3+4 K^2+2 K-2\end{array}\right),\label{970}\end{equation}

\noindent and\footnote{There seems to be a minor typo in Eq. (3.18) for $C_{\Omega,Q}$ of Ref. \cite{Lands}.  $(1+K^2)$ in the numerator should be $(1+K)^2$.}

\begin{equation} C_{\Omega,Q} \equiv T\left(\frac{\partial S}{\partial T}\right)_{\Omega,Q}=\frac{M^2 K(1+K)^2(K^2+L^2+2K)}{4A}.\label{990} \end{equation}

\par
$C_{J,Q}$ diverges if $L^2=2K$.  This may be written

\begin{equation} \alpha^2+6\alpha+4\beta=3,\label{996} \end{equation}

\noindent which gives the Davies curve, shown in Fig. 1.

\section{ACKNOWLEDGEMENTS}

\noindent I thank J. \r{A}man, B. Andresen, K. Johnston, and H. Quevedo for useful correspondence.  I also acknowledge support from the New College of Florida faculty development fund and help with the figures from Thomas Ruppeiner.

\newpage

\renewcommand{\baselinestretch}{1.0}
\normalsize

\begin{center}
\begin{tabular}{l|c|c|c|l}

\hline
\hline
System                                                                         & $n$    & $d$       &$R$ sign  &  Divergence             \\
\hline
3D Bose gas \cite{Mrug90}                                      & $2$     & $3$      &   $-$          &   $T\rightarrow 0$    \\
1D Ising ferromagnet   \cite{Rupp81,Mrug}          & $2$    & $1$        &   $-$          &  $T\rightarrow 0$     \\
Critical region  \cite{Rupp95,Rupp79,Brody95}   & $2$    & $\cdots$        &   $-$          &  critical point              \\
Mean-field theory \cite{Mrug}                                  & $2$    & $\cdots$        &   $-$          & critical point               \\
van der Waals \cite{Rupp95,Brody95}                  & $2$     & $3$      &   $-$          &  critical point              \\
Ising on Bethe lattice \cite{Dol97}                          & $2$      & $\cdots$       &   $-$          & critical point              \\
Ising on 2D random graph \cite{John,Jan02}       & $2$     & $2$      &   $-$          & critical point              \\
Spherical model \cite{John,Jan03}                         & $2$     & $\cdots$       &   $-$          & critical point               \\
Self-gravitating gas \cite{Rupp96}                          & $2$     & $3$       &   $-$          &  unclear                     \\
1D Ising antiferromagnet   \cite{Rupp81,Mrug}    & $2$    & $1$       &   $-$          &  $|R|$ small               \\
Tonks gas \cite{Rupp90B}                                        & $2$    & $1$       &   $-$          &  $|R|$ small               \\
Pure ideal gas \cite{Rupp79}                                   & $2$     & $3$       &   $0$         &  $|R|$ small              \\
Ideal paramagnet \cite{Rupp81,Mrug}                   & $2$    & $\cdots$        &   $0$         &  $|R|$ small              \\
Multicomponent ideal gas \cite{Rupp90}               & $>2$  & $3$       &   $+$         &   $|R|$ small            \\
1D Potts model \cite{John,Dol02}                            & $2$     & $1$      &  +/-            &    $T\rightarrow 0$  \\
Takahashi gas \cite{Rupp90B}                                 & $2$    & $1$       &   +/-           &  $T\rightarrow 0$    \\
Finite 1D Ising ferromagnet \cite{Brody03}              & $2$    & $1$       &  +/-            &   $T\rightarrow 0$   \\
3D Fermi gas \cite{Mrug90}                                       & $2$     & $3$      &   $+$        &  $T\rightarrow 0$    \\
3D Fermi paramagnet \cite{Kav}                              & $3$     & $3$       &   $+$        &  $T\rightarrow 0$    \\

\hline
\hline

\end{tabular}
\end{center}

\normalsize

Table I.  Thermodynamic curvature for ordinary thermodynamic systems.  I give the number of independent thermodynamic parameters $n$, spatial dimension $d$, sign of $R$, and comment on possible divergences.  In some systems $d$ is not set, and I denote this with "$\cdots$".  All signs of $R$ have here been put into the sign convention of Weinberg \cite{Weinberg}.  An indication "$|R|$ small" means $|R|$ has a value on the order of the volume of an intermolecular spacing or less.

\newpage

\noindent Figure Captions

\noindent FIG. 1.  Some characteristic curves for the Kerr-Newman black hole; see the Appendix.  The curve along which $C_{J,Q}$ diverges is the Davies curve.  $R$ diverges at the extremal limit and along curves  corresponding to a change of stability, which have diverging $C_{J,\Phi}$ and $C_{\Omega,Q}$.

\noindent FIG. 2.  The event horizon broken up into Planck area pixels.  The dark pixels are portrayed as somehow correlated.  I propose that $|R|$ measures the average number of correlated pixels.

\noindent FIG. 3.  $R(M/M_p)^2$ as a function of $J/M^2$ and $Q/M$ for $(M,J,Q)$ fluctuations.  $R$ is real, positive, and regular in the physical regime, and diverges as $T^{-1}$ at the extremal limit.

\noindent FIG. 4.  $R(M/M_p)^2$ as a function of $J/M^2$ and $Q/M$ for $(J,Q)$ fluctuations.  $R$ is real, positive, and regular in the physical regime, and diverges as $T^{-1}$ at the extremal limit.

\noindent FIG. 5.  Stable fluctuation regime for $(M, Q)$ fluctuating at fixed $J$ is indicated by + signs.  The case with $J=0$ corresponds to the Reissner-Nordstr\"{o}m black hole, which lies entirely out of the stable regime.

\noindent FIG. 6.  $R(M/M_p)^2$ as a function of $J/M^2$ and $Q/M$ for $(M,Q)$ fluctuations.  $R$ is real everywhere in the physical regime.  It is mostly positive, but there are two regimes of negative values near $Q/M=\pm1$.  $R$ diverges at both limits of stability.

\noindent FIG. 7.  Stable fluctuation regime for $(M, J)$ fluctuating at fixed $Q$ is indicated by + signs.  The case with $Q=0$ corresponds to the Kerr black hole, which lies entirely out of the stable regime.

\noindent FIG. 8.  $R(M/M_p)^2$ as a function of $J/M^2$ and $Q/M$ for $(M,J)$ fluctuations.  $R$ is real and positive everywhere in the physical regime.  $R$ diverges at both limits of stability.

\includegraphics[width=6in]{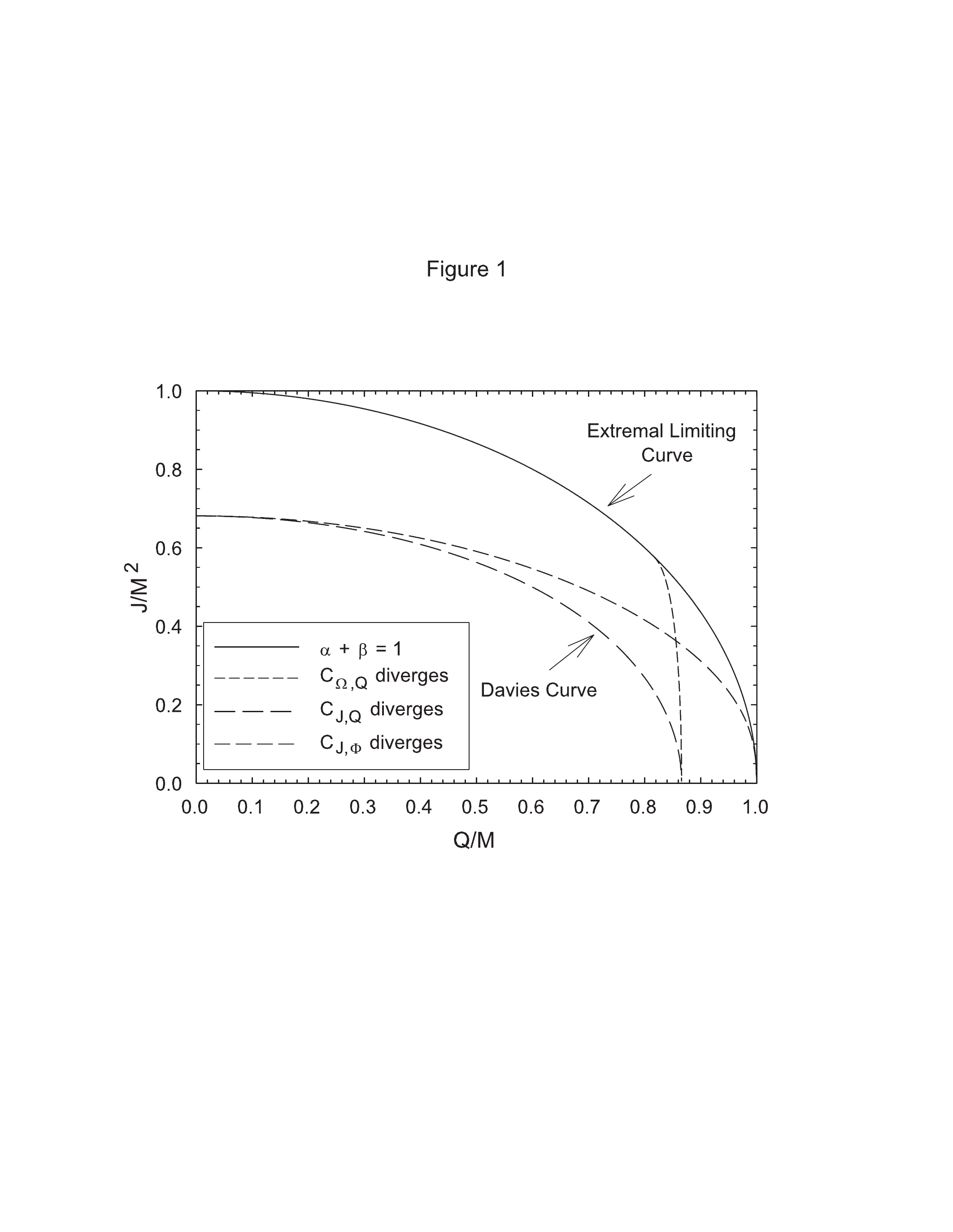}
\newpage
\includegraphics[width=6in]{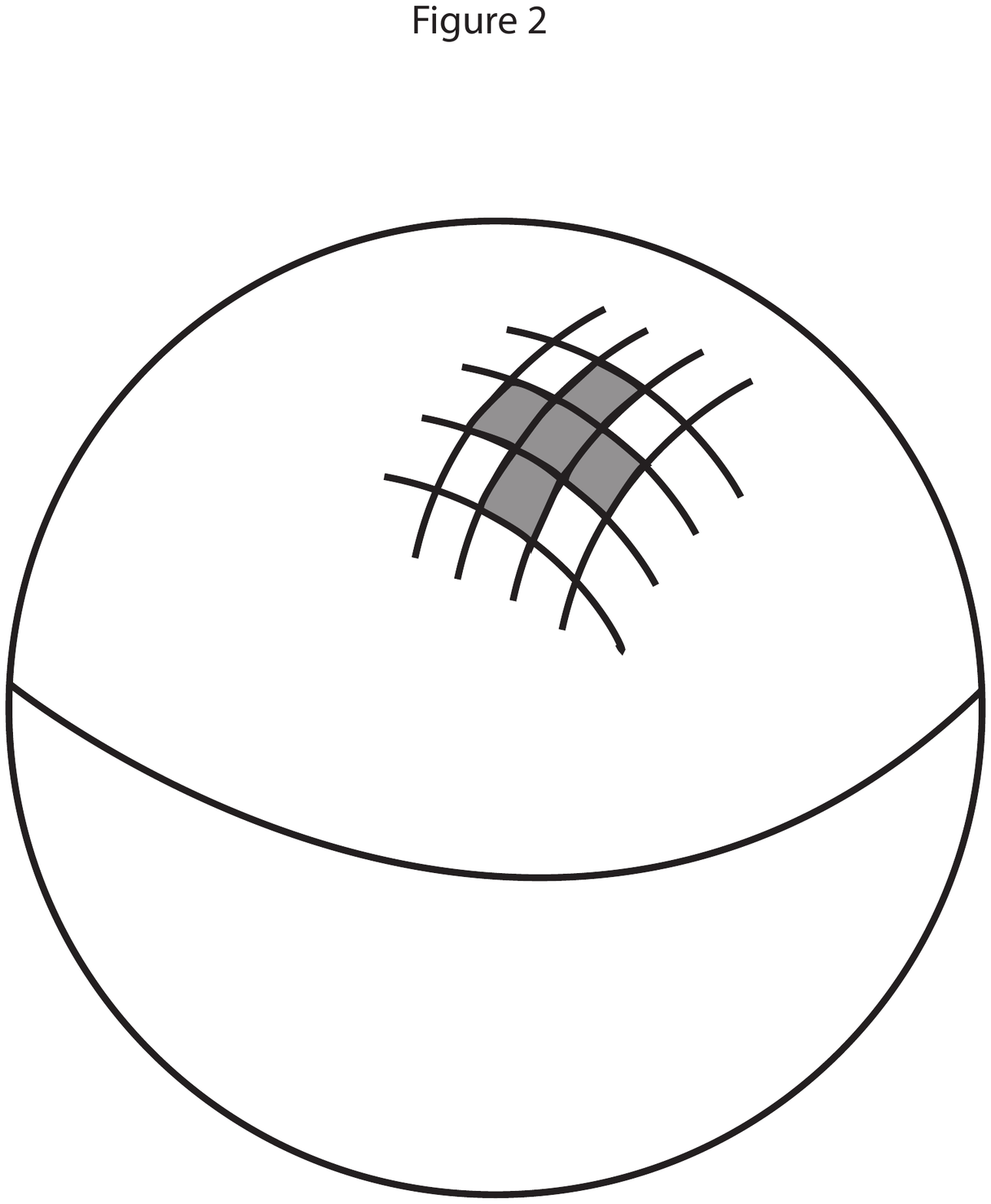}
\newpage
\includegraphics[width=6in]{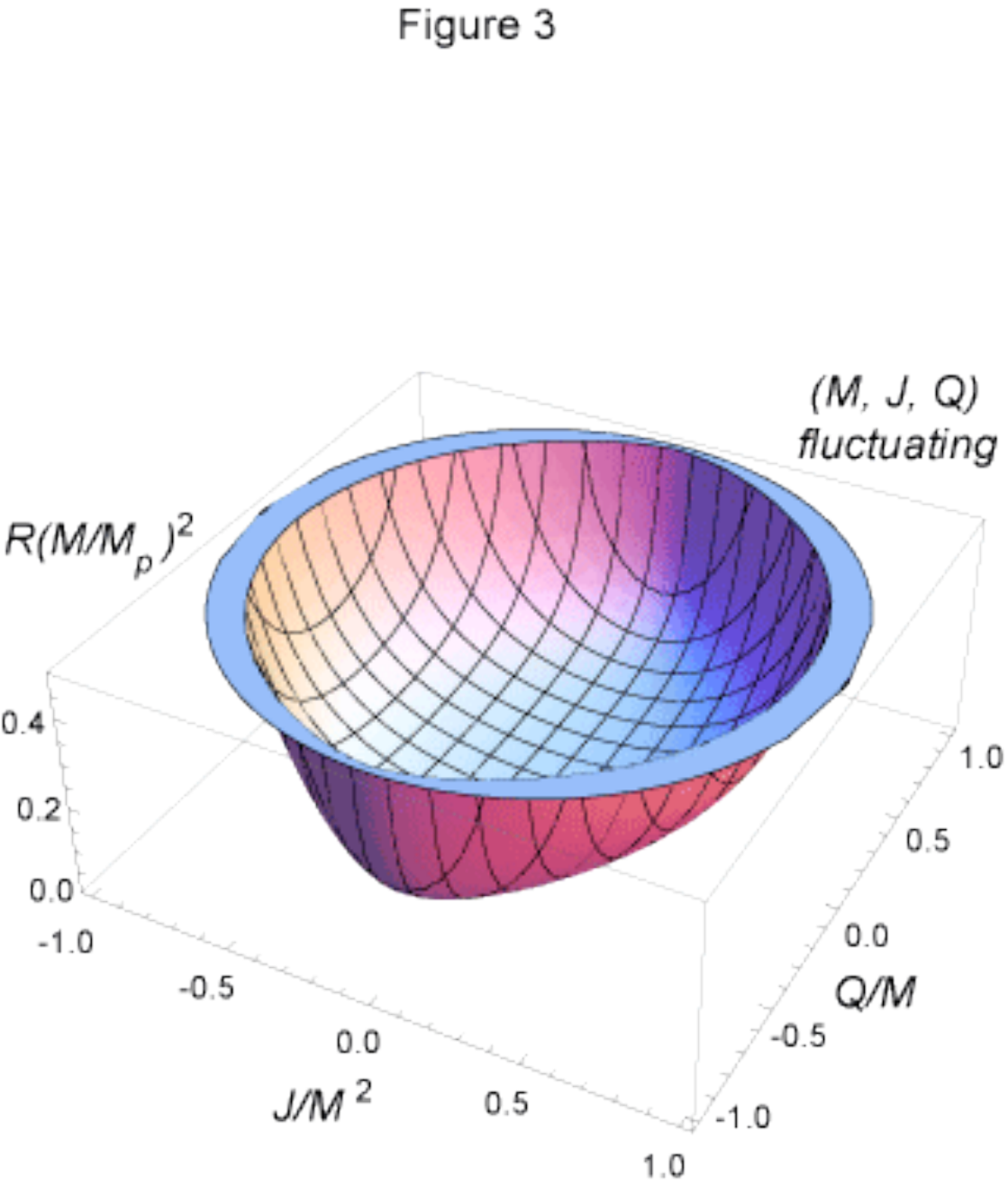}
\newpage
\includegraphics[width=6in]{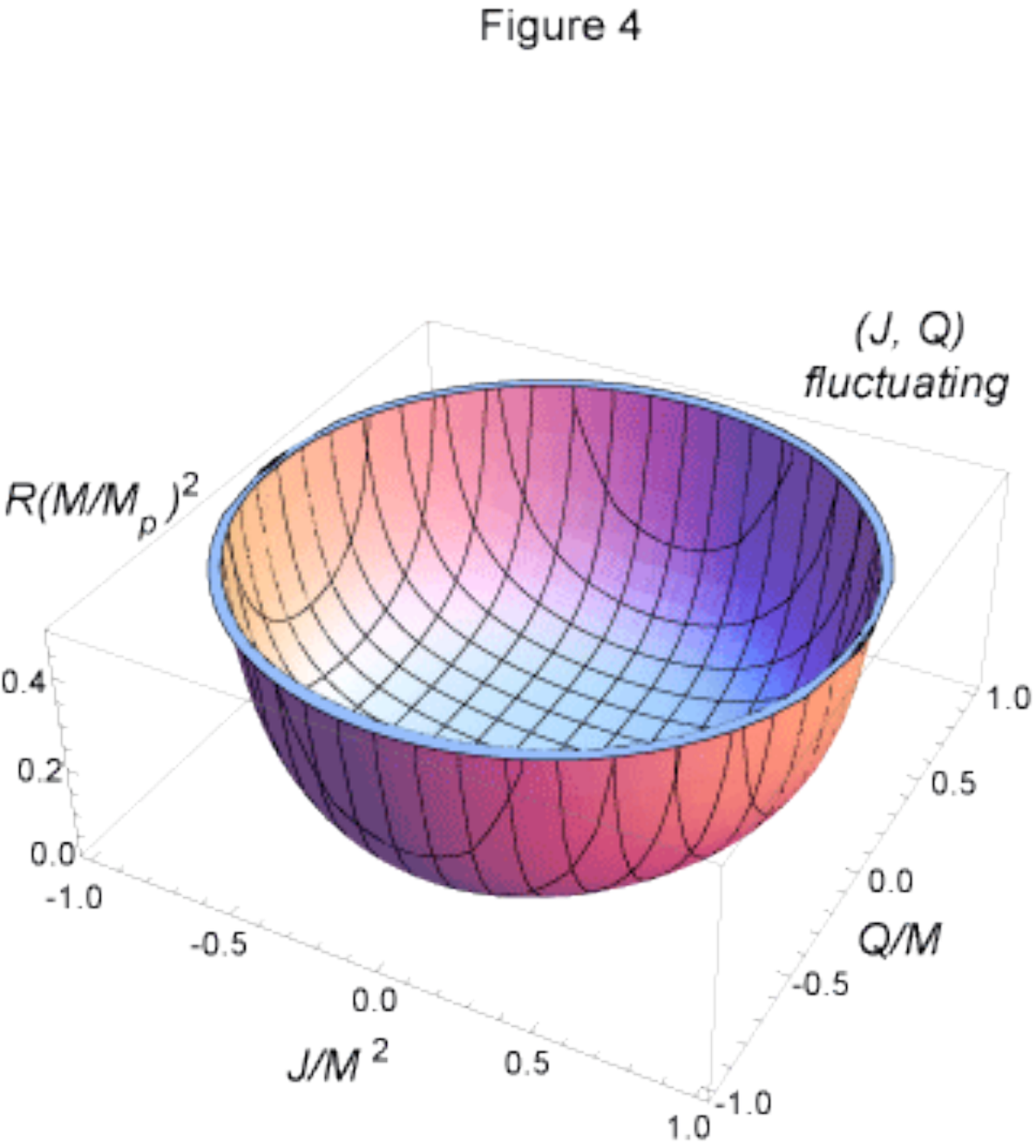}
\newpage
\includegraphics[width=6in]{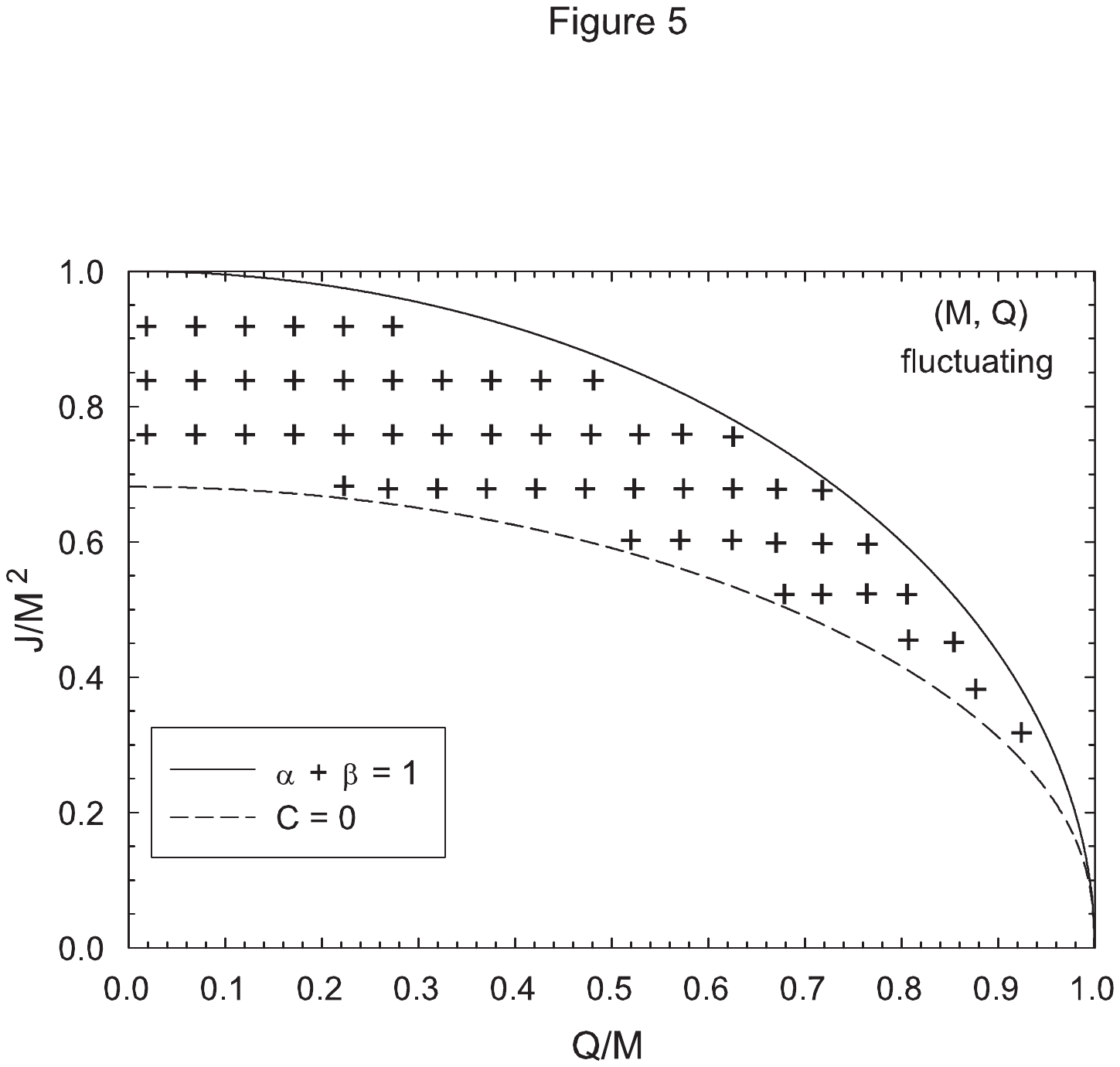}
\newpage
\includegraphics[width=6in]{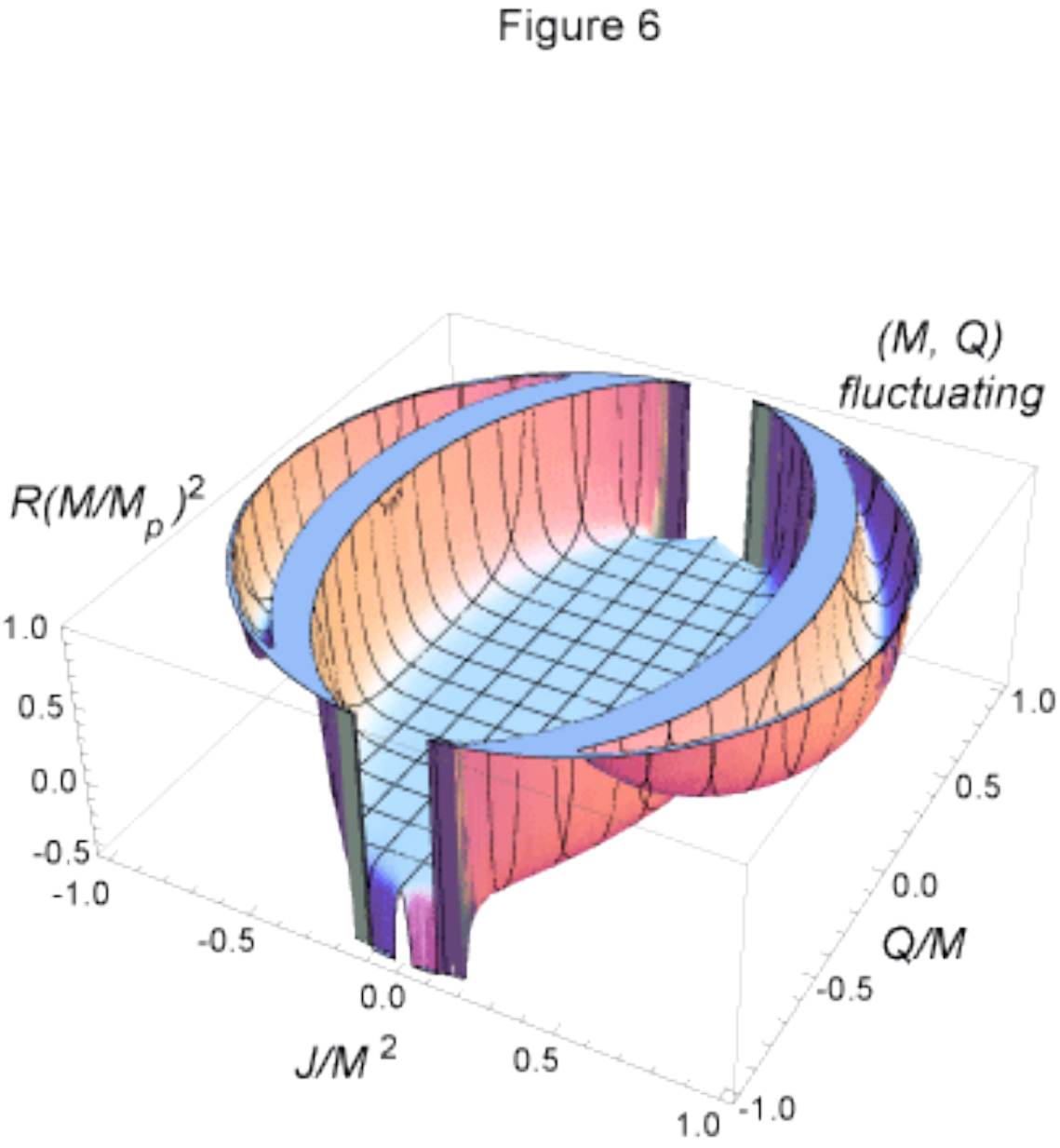}
\newpage
\includegraphics[width=6in]{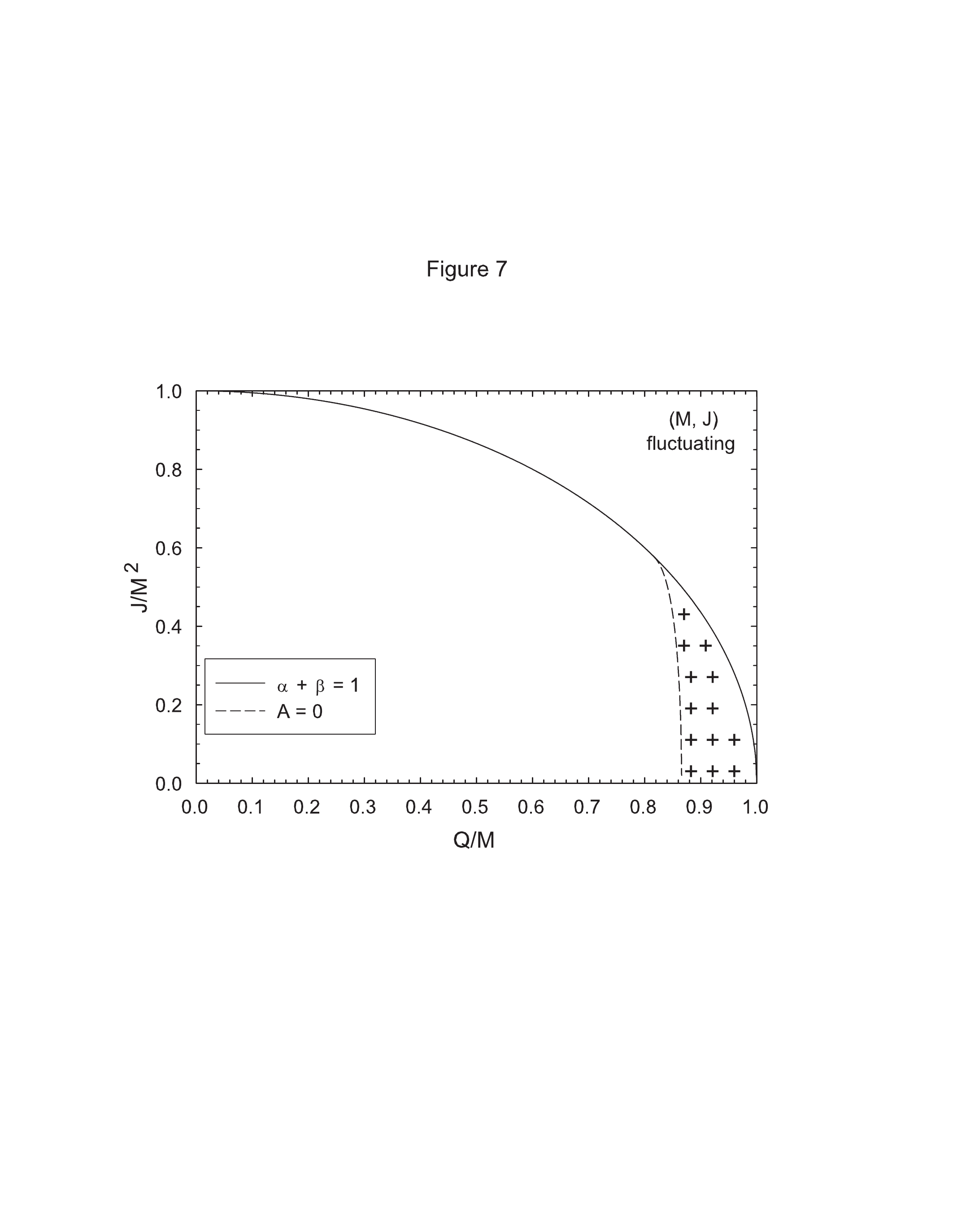}
\newpage
\includegraphics[width=6in]{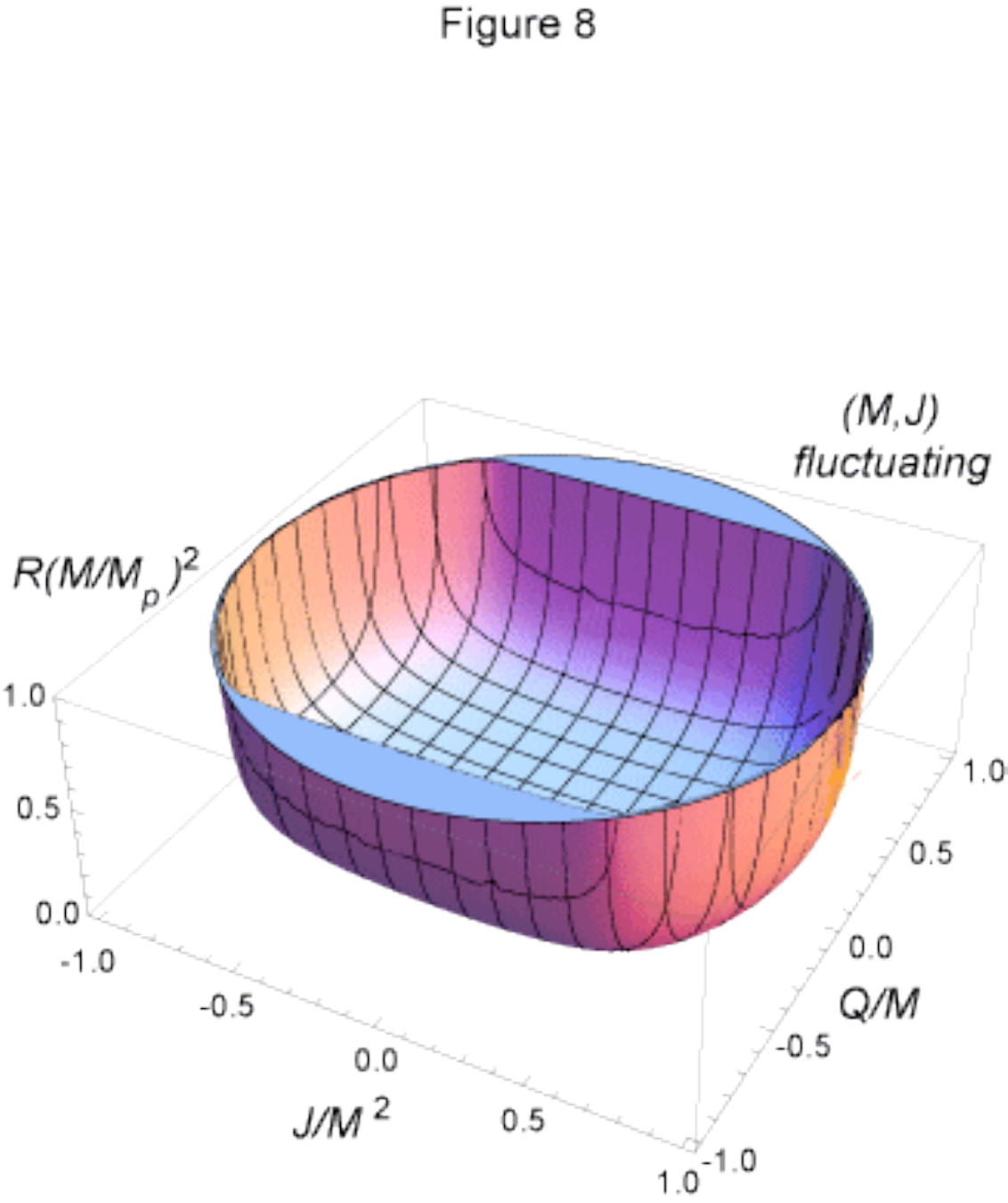}

\end{document}